\documentclass[prc,unsortedaddress,preprint,a4paper]{revtex4}
\usepackage{graphicx}
\usepackage{dcolumn}

\begin{document}

\title[Elastic and Raman ...]{Elastic and Raman scattering of 9.0 and 11.4 MeV photons   \\
from Au, Dy and In}

\author{\surname{Sylvian} Kahane}
\email{skahane@bgumail.bgu.ac.il}
\affiliation{
Physics Dept., Nuclear Research Center - Negev, \\
P. O. Box 9001, Beer-Sheva 84190, Israel}%

\author{R. Moreh}
\affiliation{
Physics Dept., Ben-Gurion Univ., Beer-Sheva, Israel
}%

\author{T. Bar-Noy}
\affiliation{
Physics Dept., Nuclear Research Center - Negev, \\
P. O. Box 9001, Beer-Sheva 84190, Israel}%

\date{\today}

\begin{abstract}
Monoenergetic photons between 8.8 and 11.4 MeV were scattered elastically and inelastically (Raman) from
natural targets of
Au, Dy and In. 15 new crosss sections were measured. Evidence is presented for a slight deformation in the $^{197}$Au
nucleus, generally believed to be spherical. It is predicted, on the basis of these measurements, that the Giant
Dipole Resonance of Dy is very similar to that of $^{160}$Gd. A narrow isolated resonance at 9.0 MeV is observed
in In.
\end{abstract}

\pacs{Valid PACS appear here}


\maketitle

\section{Introduction}
Elastic scattering of photons is interesting first of all due to the presence of Delbr\"uck scattering, named after
Max Delb\"uck, the 1969 Nobel prize recipient in biology. In a previous part of his career, as a physicist, Delbr\"uck
proposed an explanation for the forward peaked behaviour of the elastic photon scattering, as was observed by
Meitner and K\"osters~\cite{Me:33}. This is a non linear effect,  predicted by quantum electrodynamics,
 with no analogue
via the classical Maxwell equations. It is similar to the photon-photon scattering where
one of the real photons is replaced by the electrostatic potential field of a nucleus, providing a virtual photon and
enhancing the cross section. Out of the three non linear effects: photon-photon scattering, photon splitting and Delbr\"uck
scattering, only the last one was observed and thoroughly studied. However, some preliminary evidence for
photon splitting was reported in Ref.~\cite{Ak:98}.

In its lowest order,  the Born approximation, Delbr\"uck scattering consist of a diagram with
4 vertices (i.e. 4th order QCD) with a cross section proportional to $(\alpha Z)^4$.  This diagram contains a closed
electron-positron loop, i.e. the vacuum polarization, making Delbr\"uck scattering a direct
evidence of this purely quantum prediction. In higher orders, beyond the Born approximation, radiative corrections
can be added to the first order diagram. These radiative corrections are known as Coulomb corrections. Cheng and Wu \cite{CW:69}
succeded in summing up a whole class of radiative corrections, namely additional multiple photon exchange with the nucleus,
in the limit of very high energies $E_\gamma \gg mc^2$, predicting a big influence of the Coulomb corrections on the cross
section. This prediction was confirmed at 1 GeV energies by Jarlskog \textit{ et al}~\cite{Ja:73} and very recently
at 140-150 MeV, in an experiment involving a Compton backscattered laser beam, by Akhmadaliev \textit{ et al}~\cite{Ak:98}.
This last experiment used a new theoretical derivation by Lee and Milstein \cite{Le:94}, in which Delbr\"uck
scattering was expressed in terms of Green functions and the results of Cheng and Wu were recovered in a much
shorter way. It should be remarked that at these high energies the Delbr\"uck scattering is described only by an
imaginary amplitude which, via the optical theorem, is related to the absorption process of pair production.  The
vacuum polarization is described by the real amplitude which disappointingly vanishes at these energies.

Our experiment is performed at energies $E_\gamma \approx 20 mc^2$, where additional elastic scattering processes
occur. Of particular interest is the Nuclear Resonance in which internal degrees of freedom
of the nucleus are excited via the Giant Dipole Resonance ({\em GDR}). The additional processes are coherent with
Delbr\"uck scattering. The actual magnitude of the Coulomb corrections, at these energies, is not known because no
succesful calculation was performed. Evidence on the Coulomb corrections, based on experimental data, is quite
ambiguous, due to uncertainities introduced by the other coherent processes. Kahane and Moreh \cite{Ka:78} proposed
to see discrepancies between measurements and calculations in U, as evidence for Coulomb corrections. Their argument was
based on  an $\alpha Z$ dependence (Ta vs. U) and on a momentum transfer dependence (no discrepancies at small momentum
transfer).
Nolte \textit{ et al}~\cite{No:89} proposed an empirical Coulomb corrections function. They fitted such a function to
all the experiment-theory discrepancies and offered it as an universal Coulomb correction at least for the
energy interval 3 MeV $< E_\gamma <$ 12 MeV and angular interval $60^\circ < \theta < 150^\circ$. Of course the
implication is that discrepancies are caused only by neglecting Coulomb corrections. This approach did not
work out very well in the case of Bi \cite{Ru:83,Ka:86,Ka:94} where it become evident that the experiment-theory
discrepancies are mostly related to uncertainities in the {\em GDR} parameters. These parameters are obtained by
Lorentzian line fits to  ($\gamma$,tot)  measurements. In these measurements there are problems of
normalization, energy range measured (sometimes lower energies are not adequately sampled), neutron multiplicities and
so forth, resulting in quite different parameter sets from different laboratories, and usually even from different groups
in the same laboratory. These uncertainities are by far more important in generating discrepancies with the
theoretical calculations of photon scattering than any hypothetical Delbr\"uck Coulomb corrections contribution.

In the present work we assume that Delbr\"uck scattering is very well described by its Born approximation.
This assumption is consistent with the angular distribution results in Au. Therefore, all the photon scattering
data can be used to refine the {\em GDR} parameters describing the Nuclear Resonance contribution. This approach
was used before, succesfully, in the Bi case by Dale \textit{ et al}~\cite{Da:88} and by Kahane and Moreh \cite{Ka:86}.

\section{Experimental and Data Analysis Details}
\begin{figure}
  \includegraphics[scale=0.6]{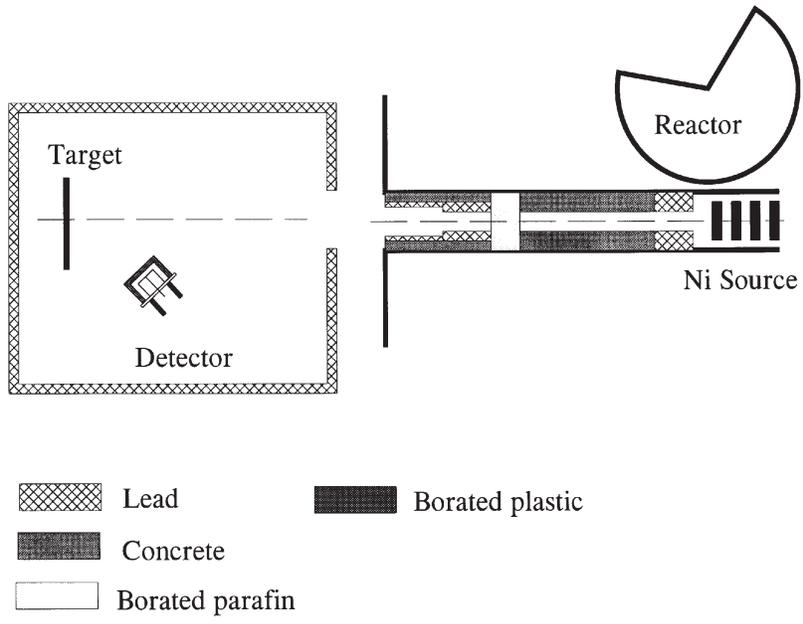}
  \caption{Schematic view of the experimental setup, not to scale.}
  \label{expset}
\end{figure}

\begin{figure}
  \includegraphics[scale=1.0]{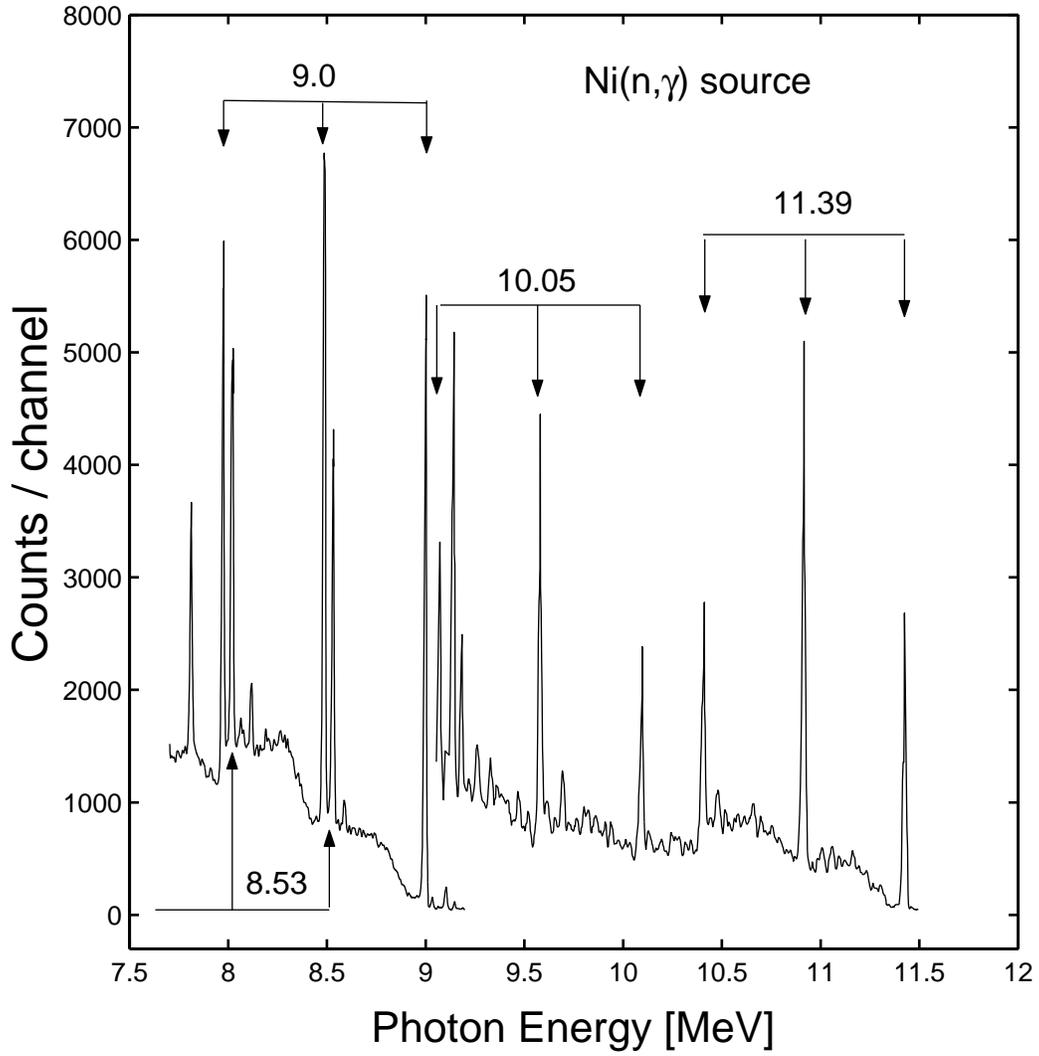}
  \caption{The spectrum of the photon beam generated by the Ni(n,$\gamma$) source in the 7.5 - 11.4 MeV
           energy range, measured after attenuating its intensity by a factor of $\approx 10^5$ using a
           lead absorber.}
  \label{dbeam}
\end{figure}

The experimental setup is described in Fig.~\ref{expset}. The source photon beam is produced by Ni(n,$\gamma$) reaction
in five separated natural Nickel metal disks, 1 Kg each, placed in a tangential beam tube, near the
core vessel of the IRR-2
nuclear reactor. The photon beam is collimated and neutron filtered along the beam tube and allowed to hit
a target placed in a lead shielded experimental chamber of $\approx$
2.0x2.0x1.5 m. Subsequently the beam is dumped into a beam catcher (not shown) designed to minimize the
backscattering toward the detector.
The Ni(n,$\gamma$) reaction produces a series of extremely sharp, well defined lines, mainly from the most
stable abundant isotope $^{58}$Ni, the most intense one is at 9.0 MeV. In distinction, the highest enegy
line at 11.4 MeV is generated \cite{Mo:72} by $^{59}$Ni, an unstable isotope with a half life of 75000
years. Our Ni source has
been under neutron bombardment for 25 years and therefore contains a sizable amount of $^{59}$Ni, produced by
neutron capture, providing a relatively strong 11.4 MeV $\gamma$ line.
Fig.~\ref{dbeam} shows the intensities of the photon beam in the energy range of interest for the present
investigation. The $\gamma$-lines appear as triplets due to the response of the 150 cm$^3$ HPGe detector
showing the
photo, first escape and double escape signals. Apart from 9.0 and 11.39 MeV there are weaker lines at
8.53 and 10.05 MeV. In studying the Au sample, another $\gamma$ source based on Cr(n,$\gamma$) reaction
was used. This source emits two intense lines at 8.88 and 9.72 MeV which were utilized in the present
measurements. The scattering angle used was 140$^{\circ} \pm$ 2$^\circ$ for the cross section
measurements and a range of 90$^\circ$ - 140$^\circ$ for angular distributions.

The yield of a scattering measurment at an angle $\theta$ is defined as:
\begin{equation}
Y_\theta = \frac{N_\theta}{t_\theta B(\theta,\phi,\mu a)}
\end{equation}
where $N_\theta$ are the net counts measured, $t_\theta$ - the measurement time and $B$ a correction for the
photon absorption in the target (see for example Ref.~\cite{Ka:94}) which depends on $\theta$,
$\phi$ - the angle of the target plane with the incoming beam direction;
$\mu$ - the linear absorption coefficient of a photon of energy $E_\gamma$ in the target material, and
$a$ - the target thickness.

In the present investigation  the cross sections were measured relative to a U standard:
\begin{equation}
\frac{d\sigma(\theta)}{d\Omega} = \frac{Y_\theta}{Y_U} \left( \frac{d\sigma(\theta)}{d\Omega} \right)_U
\frac{\Omega_U}{\Omega_\theta} \frac{n_U}{n_\theta} {\cal N}
\end{equation}
where $Y_U$ is the yield measured from the U in the same geometry as $Y_\theta$; $n_\theta$ and $n_U$ are the
number of scattering nuclei in the target and in the standard;
$\Omega_U/\Omega_\theta$ is practically equal to 1.0 under our experimental
conditions (3x3 cm targets at a distance $R_{eff}$=20 cm); ${\cal N}$ which normalizes the two 
measurements with respect to the reactor power fluctuations, obtained by monitoring the neutron flux at 
the Ni(n,$\gamma$) source position. $d\sigma(\theta)/d\Omega)_U$ is taken from Ref.~\cite{Ka:78} where 
absolute cross section measurements were performed. These cross sections were confirmed in an independent 
measurement (only at 90$^\circ$) by Rullhusen \textit{ et al}~\cite{Ru:83}.

The targets were in metallic or powder form and the quantities used were
16.11 g for Au, 26.9 g for Dy (Dy$_2$O$_3$), 61.04 g for In and 13.65 g for the U (U$_3$O$_8$)
standard.

\section{Theoretical Summary}
\subsection{Elastic scattering}
At the energies of interest for the present experiment $\approx$10 MeV, the elastic photon scattering consists of
four coherent contributions: (\textit{a}) scattering from a point charge (the $\gamma$ wavelength is much larger
than the dimensions of the nucleus) - this is the nuclear Thomson  scattering ({\em T}); ({\em b}) dipole excitation
of the internal degrees of freedom of the nucleus and subsequent return to the ground state - this is the nuclear
resonance  scattering ({\em NR}) and the nuclear excitation is known as the Giant Dipole Resonance ({\em GDR});
({\em c}) pair production and subsequent pair annihilation in the electrostatic field of the nucleus (real or
virtual i.e. vacuum polarization) - known as Delbr\"{u}ck scattering ({\em D}); ({\em d})
scattering from the electron cloud of the atom - known as Rayleigh  scattering ({\em R}).
The initial and final states in these processes are identical and therefore they are coherent.
In a linear polarization formalism the cross section is given as:
\begin{eqnarray}
\left( \frac{d\sigma}{d\Omega} \right)^{coh} & = & \frac{1}{2} r^2_0 (A_\parallel^2+A_\perp^2) \nonumber \\
&& \\
A &= &A^T + A^{NR} + A^D+A^R \nonumber
\end{eqnarray}
with $r_0$ the classical radius of the electron and the amplitudes $A$ in units of $r_0$. $A_\parallel$, $A_\perp$
are amplitudes parallel and perpendicular to the scattering plane, obtained from
$A\vec{\epsilon_1} * \vec{\epsilon_2}$, where $\vec{\epsilon_1} * \vec{\epsilon_2}$ is
the scalar product
 of the polarization vectors before and after the scattering. Perpendicular to the
scattering plane these vectors are parallel, in the scattering plane there is an angle $\theta$ between
them:
\begin{eqnarray}
A_\parallel(\theta) & =  & A \cos \theta \nonumber \\
A_\perp(\theta) & = & A
\label{parper}
\end{eqnarray}

Fano~\cite{Fa:60} had shown that in photon scattering the nucleus can receive some units of angular momentum
\textbf{ L}=0,1,2, a capability closely related to the nuclear deformation. The case \textbf{ L}=0, the scalar case,
is the coherent scattering discussed above. The vector case
\textbf{ L}=1, vanishes according to Fano, but the tensor case \textbf{ L}=2 contribute to the elastic scattering in cases
where the nuclear ground state spin $I_0 \geq 1$ and the nucleus is deformed. This contribution
to the cross section is non coherent because the final state differes by 2 units of angular momentum compared
with the initial state~\cite{Ha:70}; its form in the modified simple rotor model \cite{TM:77} is:
\begin{widetext}
\begin{equation}
\left( \frac{d\sigma}{d\Omega} \right)^{incoh} = r_0^2 (I_0K_020|I_0K_0)^2\left| P\times A^{NR}_1-A^{NR}_2 \right|^2
\frac{13+\cos^2 \theta}{40}
\label{incoh}
\end{equation}
\end{widetext}
$K_0$ is the nuclear spin projection on the nuclear symmetry axis and $P$ is given below.
The $A^{NR}$ amplitude (at $\theta=0$) is obtained from the
Lorentzian parameters of the {\em GDR} (the central energy $E$, the width $\Gamma$ and the maximum cross section
$\sigma$ at $E$) and the photon energy $E_\gamma$ \cite{Ka:74} (in units of $r_0$):
\begin{widetext}
\begin{equation}
A^{NR}(E_\gamma) = \left( \frac{\alpha}{4\pi} \right) \left( \frac{E_\gamma}{mc^2} \right) \left(
\frac{\sigma}{r_0^2} \right) \Gamma E_\gamma \frac{E^2-E_\gamma^2+i \Gamma E_\gamma}
{(E^2-E_\gamma^2)^2+\Gamma^2 E_\gamma^2}
\label{eqNR}
\end{equation}
\end{widetext}
For a deformed nucleus the {\em GDR} is split in two peaks with two sets of Lorentzian parameters: $E_i$,
$\Gamma_i$, $\sigma_i$ ($i$=1,2) and hence two amplitudes $A^{NR}_i$; where the coherent amplitude is
$A^{NR}=A^{NR}_1+A^{NR}_2$ and the factor $P$ in Eq.~(\ref{incoh}) is the ratio
$\sigma_2 \Gamma_2 / \sigma_1 \Gamma_1$.
For a non deformed nucleus, or a $I_0<1$ nucleus, the incoherent contribution to the elastic scattering vanishes.

The Thomson amplitude is given \cite{GGT:54}, for $E_\gamma=0$ and $\theta=0$, as $A^T=-Z^2m/M$ where $m$
is the electron mass and $M$ the nuclear mass. In principle for $E_\gamma > 0$ there are additional terms
\cite{Ru:83} based on the form factor of the nuclear charge distribution and exchange terms. For our energies
these corrections are negligible.

Delbr\"uck scattering amplitudes were calculated numerically by Kahane~\cite{Ka:92} and by Bar-Noy and
Kahane~\cite{Tu:77}, in the Born approximation, using the formalisms of
Papatzacos and Mork~\cite{Pa:75} and De Tollis \textit{ et al}~\cite{DT:71}.

Rayleigh scattering  was calculated in its first order by a second order $S$ matrix formalism by Kissel
\textit{ et al}~\cite{Ki:80}. Unfortunately numerical results exist only for lower energies $<$ 2.754 MeV. Besides
the exact $S$ matrix calculations, the most popular approximation to Rayleigh scattering is the Modified Relativistic
Form Factor {\em MRFF}~\cite{Br:55} which depends only on the momentum transfer. This approximation is
not so good beyond momentum transfers $q\approx$10\AA$^{-1}$. In our experiment at 9.0 MeV and 140$^\circ$
$q\approx$682\AA$^{-1}$.

Table~\ref{Table1} summarizes the amplitudes for the elastic scattering for Au at 9.0 MeV and 140$^\circ$.
The {\em R} amplitudes were taken from the internet site of Ref.~\cite{Ki:80} in the {\em MRFF} approximation
(file: \verb+079_cs0sl_mf+); {\em D} amplitudes from Ref.~\cite{Ka:92}; {\em NR} amplitudes from 
Eq.~(\ref{eqNR})
with the {\em GDR} parameters of Fultz \textit{ et al}~\cite{Fu:62}.
It seems that the {\em R} amplitudes are very small compared to the other contributions. The interference
terms contributed by the {\em R} amplitudes have only a small influence, $\approx$0.1\%, on the scattering cross
section. Thus, the {\em R} scattering amplitudes were neglected.
The same conclusion was
reached by us before, on the basis of the {\em R} calculations of Florescu and Gavrila~\cite{Fl:76}. These
calculations are exact in the sense that they employ second order $S$ matrix but not realistic in the sense that
only the $K$-shell electrons are calculated in a pure Coulomb field (enabling an analytic evaluation).
On the contrary, the {\em MRFF} is not exact (momentum transfer far beyond the range of applicability), but  is
realistic with all the electrons included and employing a self consistent atomic potential. The conclusion is
equally valid for the other energy and targets used. At 11.4 MeV the {\em R} amplitude decreases because with increasing
energy the {\em R} scattering becomes more forwardly peaked. For Dy and In, of lower $Z$, the {\em R}
amplitude decreases because of its strong $Z^2$ dependence.

\begin{figure}
  \includegraphics[scale=1.0]{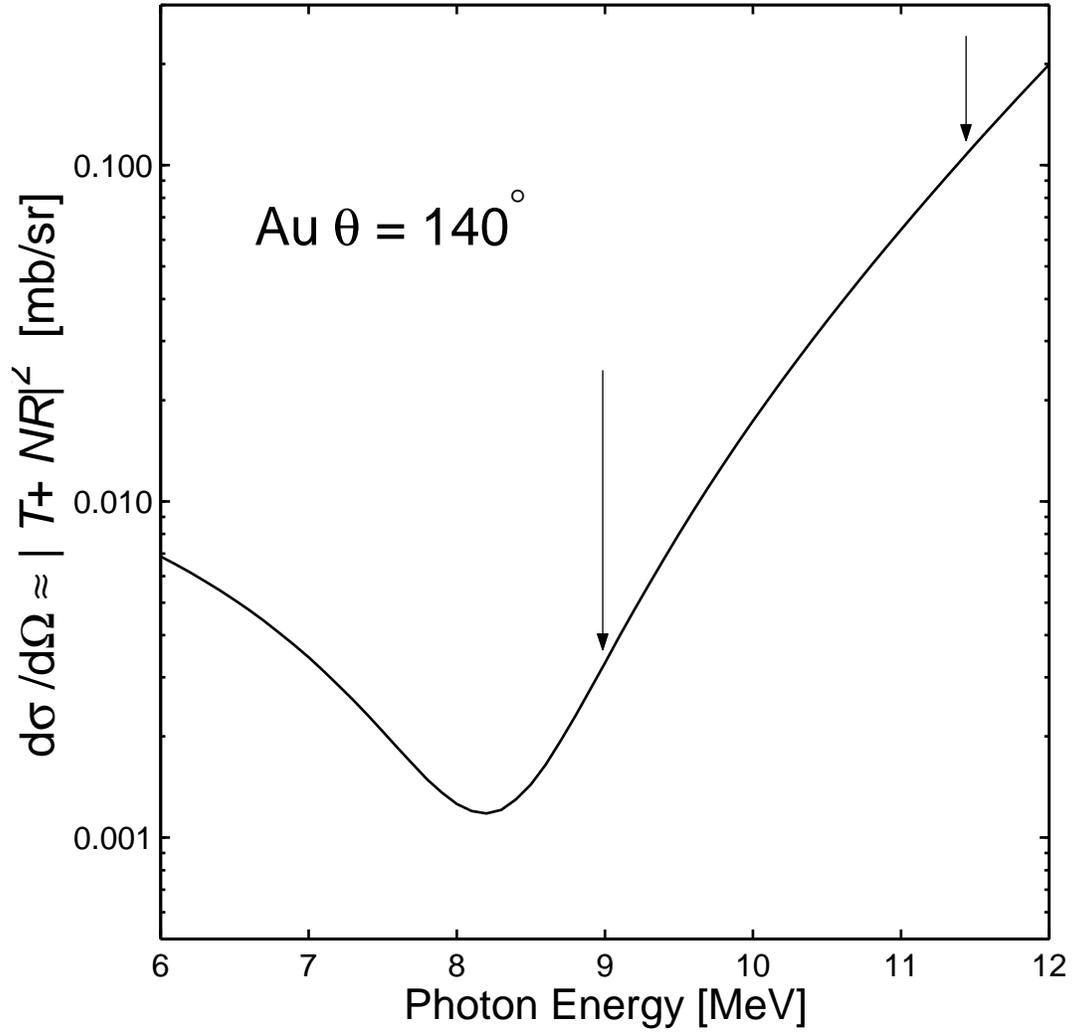}
  \caption{Destructive interference between {\em T} and {\em NR} contributions in Au. The arrows point to the exact
locations of the 9.0 MeV and 11.4 MeV.}
  \label{Auinter}
\end{figure}

A destructive interference effect, predicted by \cite{GGT:54}, occurs between {\em T} and {\em NR}. This is
illustrated in Fig.~\ref{Auinter} where the scattering cross section is calculated versus energy.
The destructive interference is evident as it lowers the cross section in the 8-9 MeV range, and
is expected to show up in the experimental measurements as well, even if somehow masked by the additional {\em D}
contribution.

The present elastic photon scattering
results are used for deducing a best set of {\em GDR} parameters because of the high sensitivity of the data.
A summary of all the {\em GDR} parameters tested is shown in
Table ~\ref{Table2}.
There are no measured {\em GDR} parameters for Dy, most probably because in natural form it contains 7 different stable
isotopes out of which 5 are even-even nuclei. We tried to analyze the results in terms of $^{165}$Ho and  $^{160}$Gd
parameters, both being close to the most abundant $^{164}$Dy isotope.

\subsection{Raman scattering}
Deformed nuclei are characterized by rotational spectra with a rotational band including the ground state and
the low lying excited states. The photon tensor scattering gives rise to nonelastic contributions involving decay

of the {\em GDR} to  these low lying  rotational states of the nucleus. These contributions are known as nuclear
Raman scattering in analogy to the molecular Raman scattering. The cross section is given in total analogy with
Eq.~(\ref{incoh}):
\begin{widetext}
\begin{equation}
\left( \frac{d\sigma}{d\Omega} \right)^{Raman} = r_0^2 (I_0K_020|I_fK_0)^2\left| P \times A^{NR}_1-A^{NR}_2 \right|^2
\frac{13+\cos^2 \theta}{40}
\label{ramanXS}
\end{equation}
\end{widetext}

the final state spin $I_f$ refers to the level spin including the ground state spin; the strength of the
tensorial part is split between the ground state and the excited states according to the CG coefficient.

\section{Results and discussion}
Fig.~\ref{Spectra} presents the photon scattering spectra measured from the three targets. The accumulation times
were 192h for Au,
97h for Dy and 66h for In. For Au and Dy a stronger signal is observed at 11.4 MeV compared with 9.0 MeV as expected
from Fig.~\ref{Auinter}. In gives a much stronger signal at 9.0 MeV and at other lower energies, compared
with 11.4 MeV. This is due to scattering from an isolated resonance level in In and is reminiscent of our former
investigation \cite{Ka:94} of Pb isotopes where strong departure
from the smooth behavior of a Lorentzian {\em GDR} was observed. In Dy, which is a deformed nucleus, also the inelastic
Raman scattering is clearly observed. The measured cross sections are presented in Table~\ref{Table3}.

\subsection{Au}
We begin the description of Au results with the angular distributions because of the implications  of these
results on the accuracy of the {\em D} amplitudes.

\subsubsection{Angular Distributions}
\begin{figure}
  \includegraphics[scale=1.0]{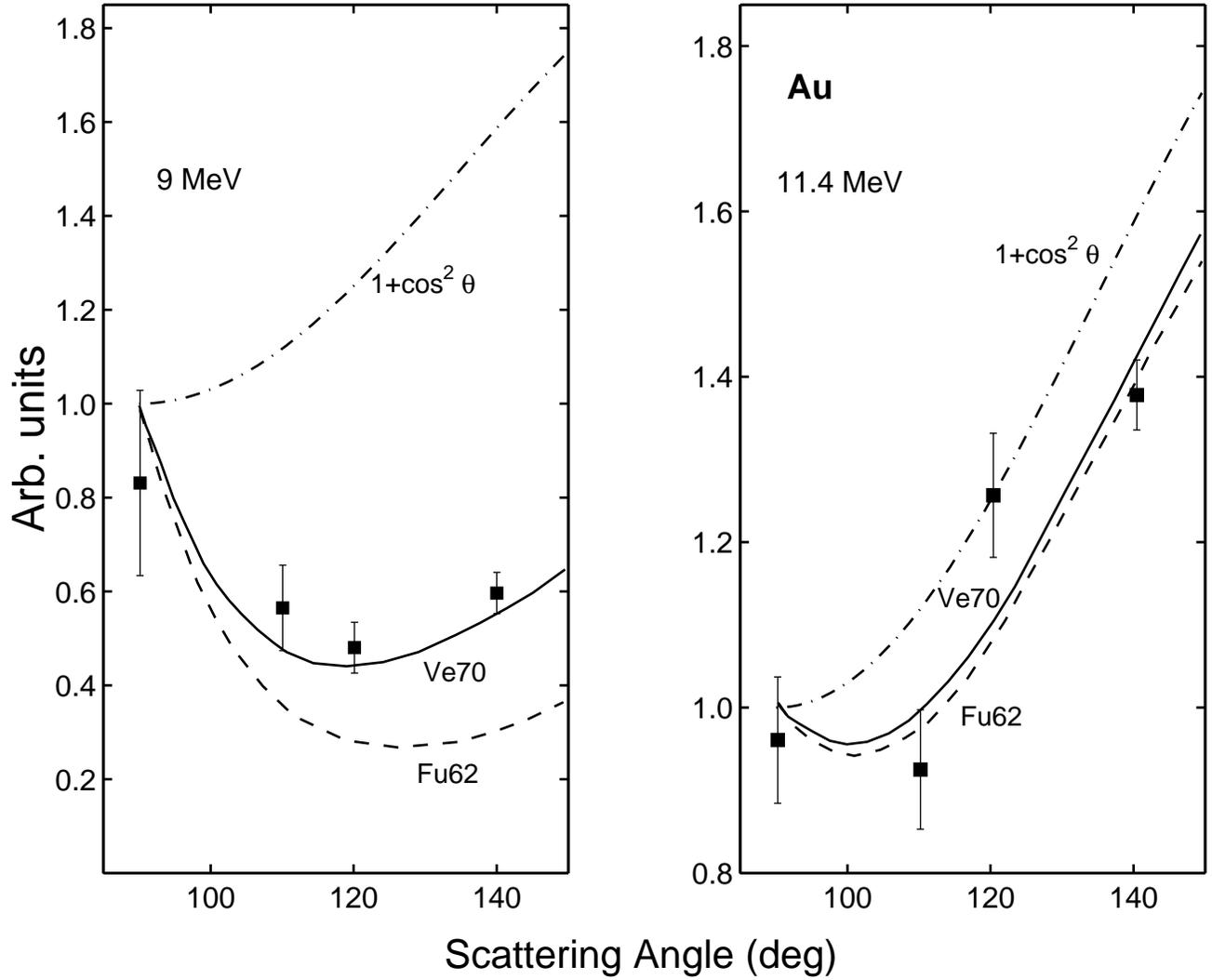}
  \caption{Measured angular distributions in Au. Calculations based on two sets of {\em GDR} parameters.
The $1+\cos^2 \theta$ dependence of {\em T} + {\em NR} is also shown.}
  \label{angd}
\end{figure}

The measured angular distributions at 9.0 and 11.4 MeV are presented in Fig.~\ref{angd}.
Calculations based on {\em T}, {\em D} - in the first Born approximation, and {\em NR} based on two sets of
{\em GDR} parameters from Table~\ref{Table2} are shown. Should {\em D} be negligible, the {\em T}
and {\em NR} would reveal an angular dependence of the form 1+$\cos^2 \theta$  (shown in Fig.~\ref{angd}).
At 11.4 MeV the measured angular distribution {\em resembles} quite closely a 1+$\cos^2 \theta$ behavior.
The explanation rests on the fact that the {\em NR} contribution becomes dominant at energies approaching
the {\em GDR} peak at $\approx$14 MeV, the contribution
of {\em D} decreases, and therefore the angular distribution {\em approaches} 1+$\cos^2 \theta$. Conversely, at
9.0 MeV the contribution of {\em T+NR} is low because of their destructive interference, {\em D} is strong, causing
a large departure from 1+$\cos^2 \theta$. One remark concerning the importance of the Coulomb corrections to the
{\em D} contribution is in order. At 11.4 MeV their contribution is not important because of the
dominance of the {\em NR} component. At 9.0 MeV, where {\em D} is dominant,
the good existing agreement between the measurement and  the calculations implies that the Coulomb
corrections are not important, at these energies, momentum transfers, and $\alpha Z < 0.58$, appropriate
for Au or lighter nuclei.

\subsubsection{Cross sections}
\begin{figure}
  \includegraphics[scale=1.0]{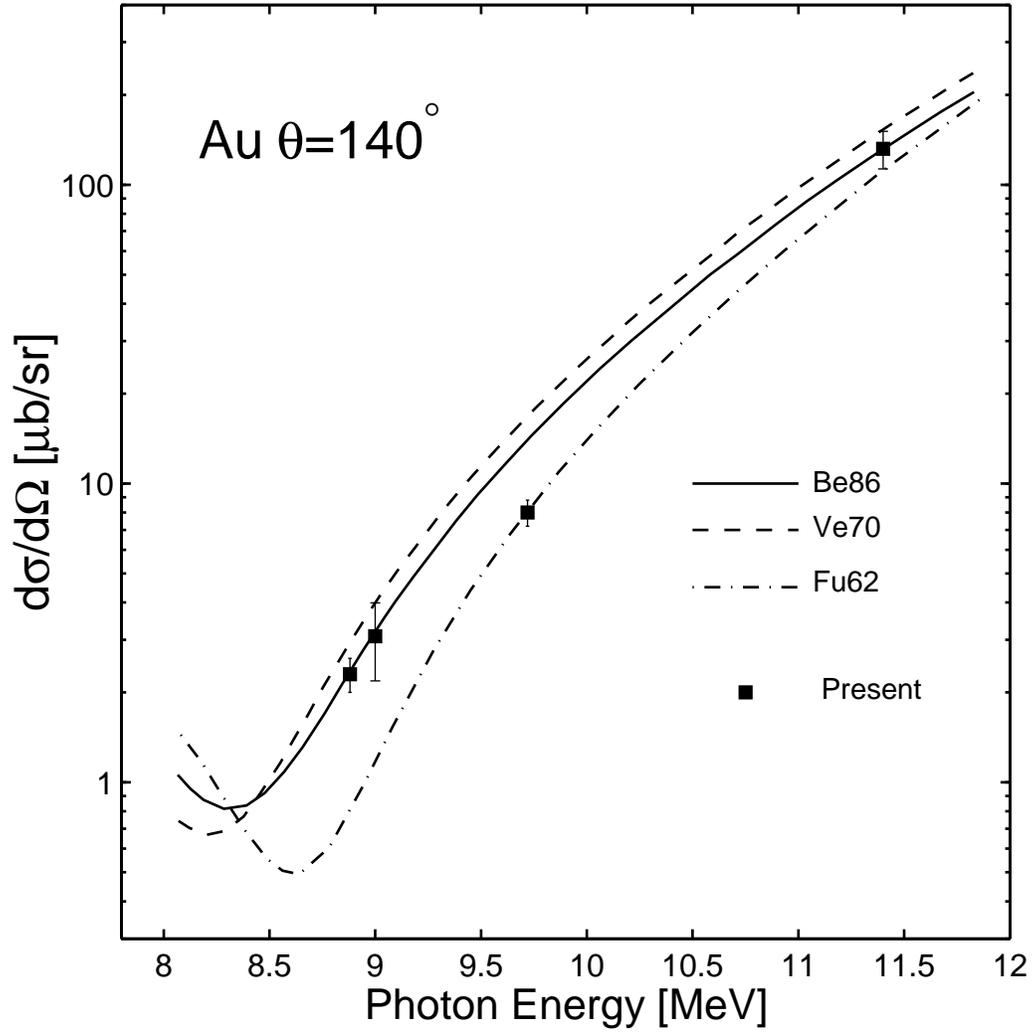}
  \caption{Measured cross sections in Au. Calculations based on three sets of {\em GDR} parameters.}
  \label{XSfig}
\end{figure}

Present results are shown in Fig.~\ref{XSfig}. They include two measurements at 8.88 MeV and 9.72 MeV
obtained with a Cr(n,$\gamma$) photon source. Three calculations based on different Au {\em GDR}
parameters from Table~\ref{Table2} are also shown. The measured value at 9.72 MeV seems to be
too low. The older parameters Fu62 clearly do not reproduce the data
correctly, neither the cross sections nor the angular distributions at 9.0 MeV. This set has a too low value of
$\Gamma$, probably due to an incomplete range of energies measured, coming too low in the scattering cross
sections at the energies near 9.0 MeV. Our measurements clearly prefer the {\em GDR} parameters from Be86~\cite{Be:86}.
This set is close to the one of Ve70~\cite{Ve:70}, having almost equal values of $\sigma \Gamma$ being 2389 vs. 2494
(in units of mb$\cdot$MeV), which is a measure of the {\em GDR}

strength. The $\Gamma$ of Be86~\cite{Be:86} is largest accounting well for the wings of the {\em GDR}.
The parameters of So73~\cite{So:73} (not
shown) have a narrow $\Gamma$ and higher strength $\sigma \Gamma$=2655 mb$\cdot$MeV.

\subsubsection{Possible deformation in $^{197}$Au}

\begin{figure}
  \includegraphics[scale=1.0]{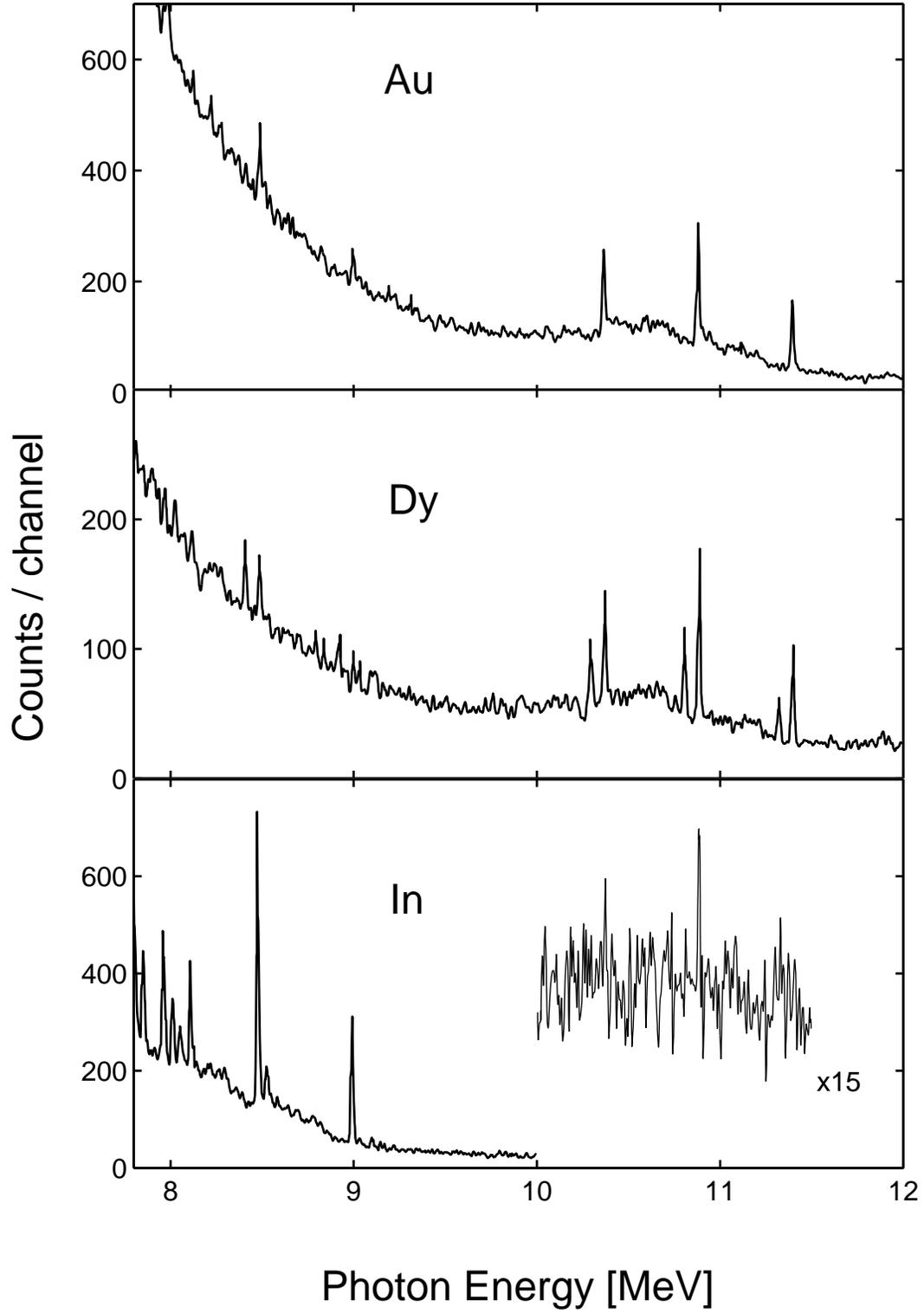}
  \caption{Measured spectra from the three targets of Au, Dy and In at $\theta=140^\circ$.}
  \label{Spectra}
\end{figure}

\begin{figure}
  \includegraphics[scale=0.8]{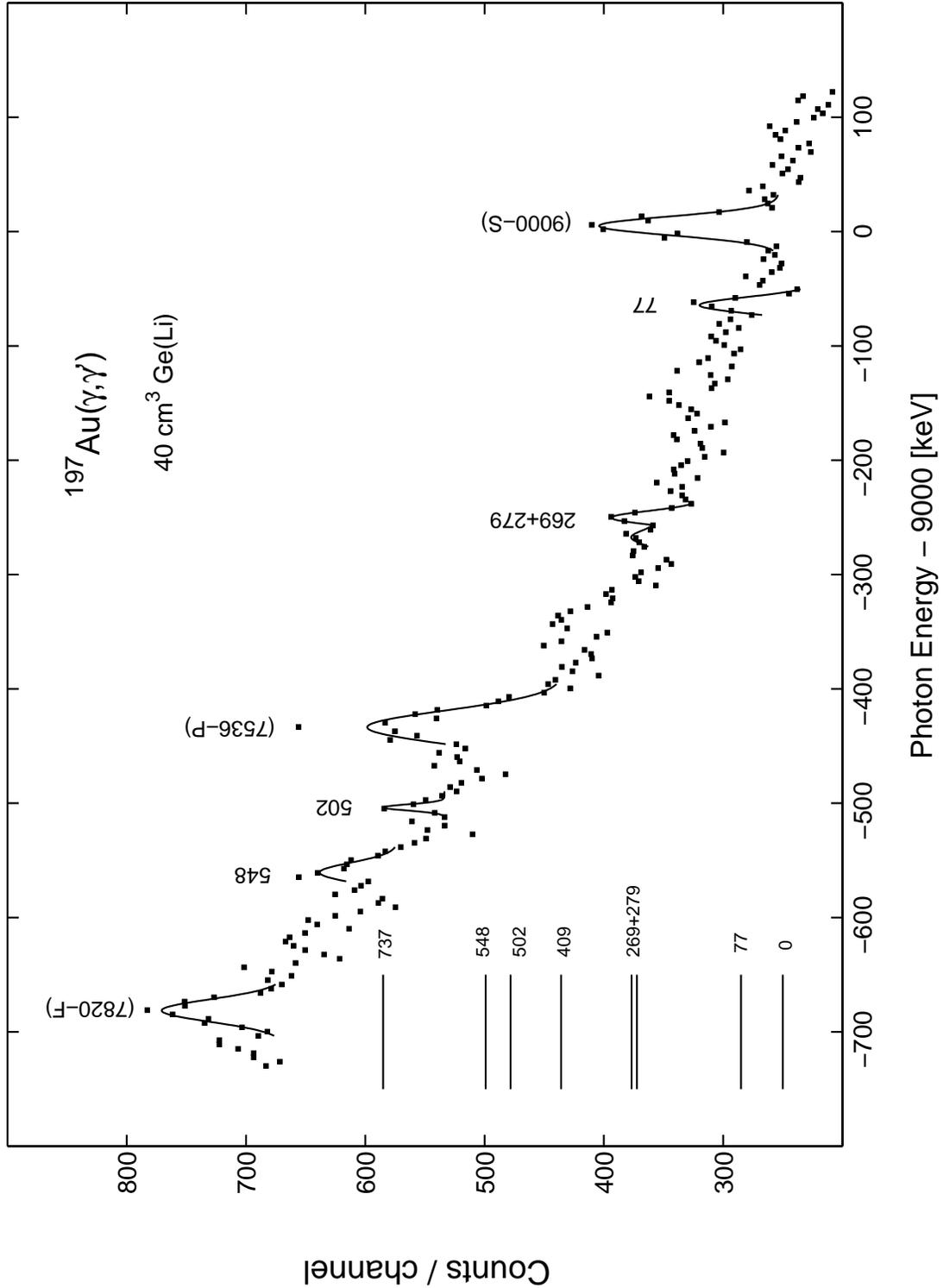}
  \caption{Inelastic transitions observed in photon scattering from Au. Abscissa is the energy difference in respect
to 9.0 MeV. Peaks with their energy noted without parantheses are inelastic transitions to the low lying states of
$^{197}$Au. Elastic transitions are noted with the energy of the incoming photon in parantheses (P -
photopeak, F - first escape, S - second escape). The inset shows a level scheme of $^{197}$Au with 
levels taken from Table of Isotopes.}
  \label{ramanAu}
\end{figure}

The $^{197}$Au is usually assumed to be spherical with a {\em GDR} having a single peak. This will imply an
absence of Raman scattering signals. The experimental result at 11.4 MeV (Fig.~\ref{Spectra}), performed using
a small target of only 16 g, seem to agree with the above expectation. At 9 MeV however, the spectrum
(Fig.~\ref{ramanAu}) reveals several inelastic transitions leading to the levels at 77, 269, 279, 502 and
548 keV in $^{197}$Au. In this later measurement, a bigger target, a more intense beam, much longer running
time but a smaller detector were used. It should be reminded that the 9 MeV line is the most intense line of
the $\gamma$ source  and is 25 times stronger that the 11.4 MeV line. A transition to the 
$\frac{11}{2}^-$ 409 keV level is forbidden by its spin and parity.

\begin{figure}
  \includegraphics[scale=1.0]{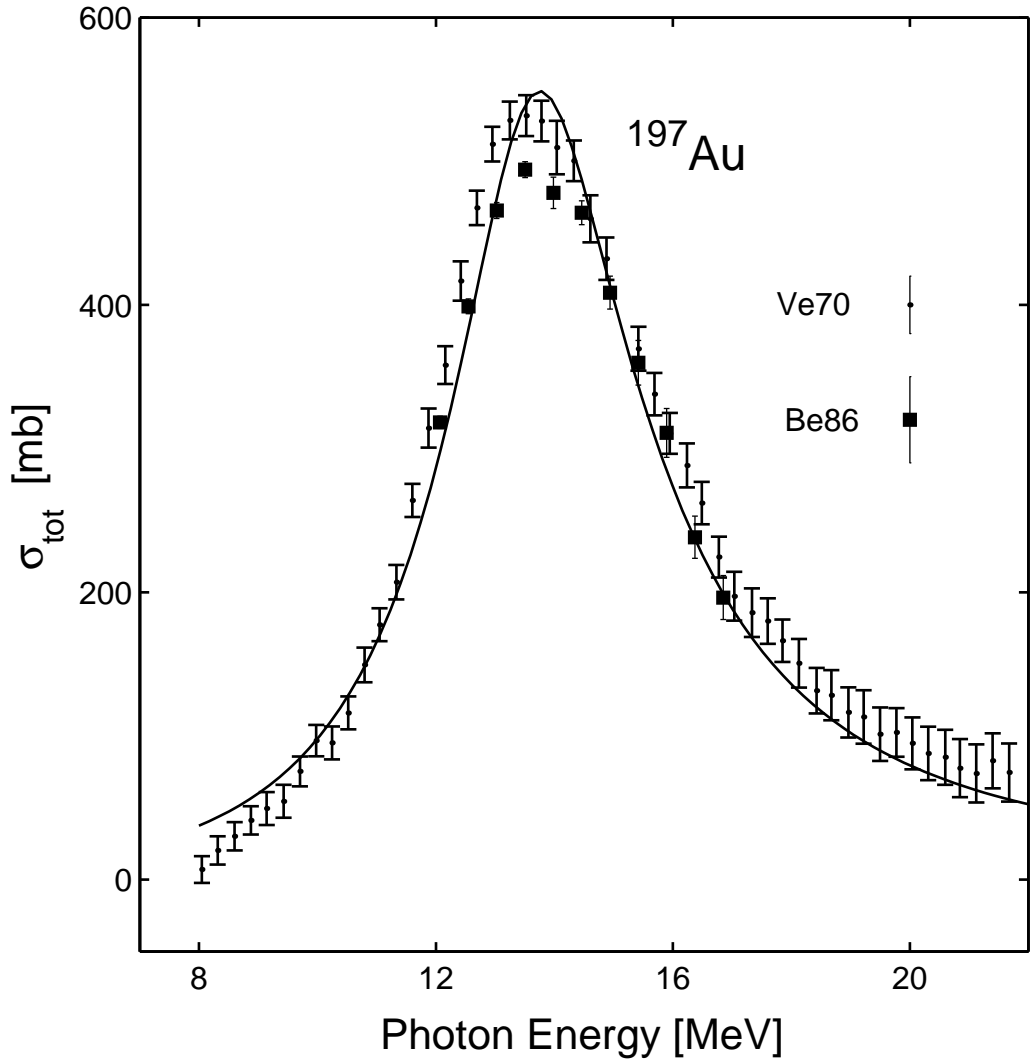}
  \caption{Two Lorentzian fit to the $^{197}$Au($\gamma$,tot) cross sections of Ve70 and Be86.}
  \label{Augn}
\end{figure}

These results came as a surprise because neither the dynamic collective model (DCM) \cite{Ar:69} nor the
simple rotator model (SRM) \cite{Ful:62} predict a nonzero Raman scattering in a nondeformed nucleus. A
tentative explanation will be that $^{197}$Au posseses a very slight deformation not easily observed.
In Fig.~\ref{Augn} composite ($\gamma$,tot) data of Ve70 and Be86 is fitted (manual adjustment) with a 
two Lorentzian line
constrained to a very small peak energy difference of 200 keV. The Ve70 ($\gamma$,tot) data were
obtained directly from Ref.~\cite{Di:88}; the Be86 data were reconstructed from the ($\gamma$,n)
+ ($\gamma$,n+p) + ($\gamma$,2n) components taken from the EXFOR system~\cite{EX:01}. 
The resulting fitting parameters are included in Table~\ref{Table2}.

The extracted experimental Raman cross sections are presented in Table~\ref{Table4}. There are large errors
because the cross sections are small and the statistical quality of the spectrum is not good.  The 
low-lying levels in Au can be arranged in two rotational-like bands: \textit{i)} a ground state band
$ \textrm{0}\ \left( \frac{3}{2} \right) \rightarrow \textrm{279} \left( \frac{5}{2} \right) 
\rightarrow  \textrm{548}\ \left( \frac{7}{2} \right) $ and \textit{ii)} a side band
$ \textrm{77} \left(
\frac{1}{2} \right ) \rightarrow \textrm{269}\ \left( \frac{3}{2} \right) \rightarrow \textrm{502}\
\left( \frac{5}{2} \right) \rightarrow \textrm{737}\ \left( \frac{7}{2} \right) $; each one fitted
nicely by an expression of the form $E(K,I)=E_K+AI(I+1)+BI^2(I+1)^2$ \cite{Bo:75} with similar
values for the coefficients $A$ and $B$. $K$ is given by the spin $I$ of the band head \cite{Bo:75}. 
In a given band the tensor cross section is shared between different transitions according to the 
CG coefficients in Eq.~(\ref{ramanXS}) (sum of their squares is 1). Only in the {\em DCM} one can 
calculate how the cross section is shared between different bands.
Also presented in Table~\ref{Table4} are Raman cross sections calculations based on Eq.~(\ref{ramanXS}) 
({\em SRM}) and the above two Lorentzian fit parameters. Because there is no division of the inelastic
cross section strength between the $K_0=\frac{1}{2}$ and $K_0=\frac{3}{2}$ bands in {\em SRM}, 
these calculations provides only an upper limit (they assume that the full Raman strength is 
feeding the band).
While the calculated cross sections are consistently somewhat higher than the experiment, there is
quantitative agreement within one standard deviation for the $K_0=\frac{1}{2}$ calculations. The 
$K_0=\frac{3}{2}$ calculation overestimates the experimental results notably at 279 keV.
Also, the calculated incoherent contribution to the elastic transition is too large. It seems,
therefore, that the $K_0=\frac12$ band receives a greater share of the Raman strength compared
with the $K_0=\frac32$ band.

The calculated Raman cross section for the 77 keV transition at $E_\gamma$=11.4 MeV is 8.5~$\mu$b/sr,
a factor 15 lower than the elastic cross section. The signal to noise ratio for the elastic peak at
11.4 MeV (first escape) is 0.25 (Au spectrum from Fig.~\ref{Spectra}). This implies an expected signal to 
noise ratio for the Raman peak of only 0.015, i.e. only 1.5\% over the background while the background
itself has a statistical uncertainity of $\approx$2\% - 3\%.
It explains why the Raman signal was not detected at $E_\gamma$=11.4 MeV.

\subsection{Dy}
\begin{figure}
  \includegraphics[scale=1.0]{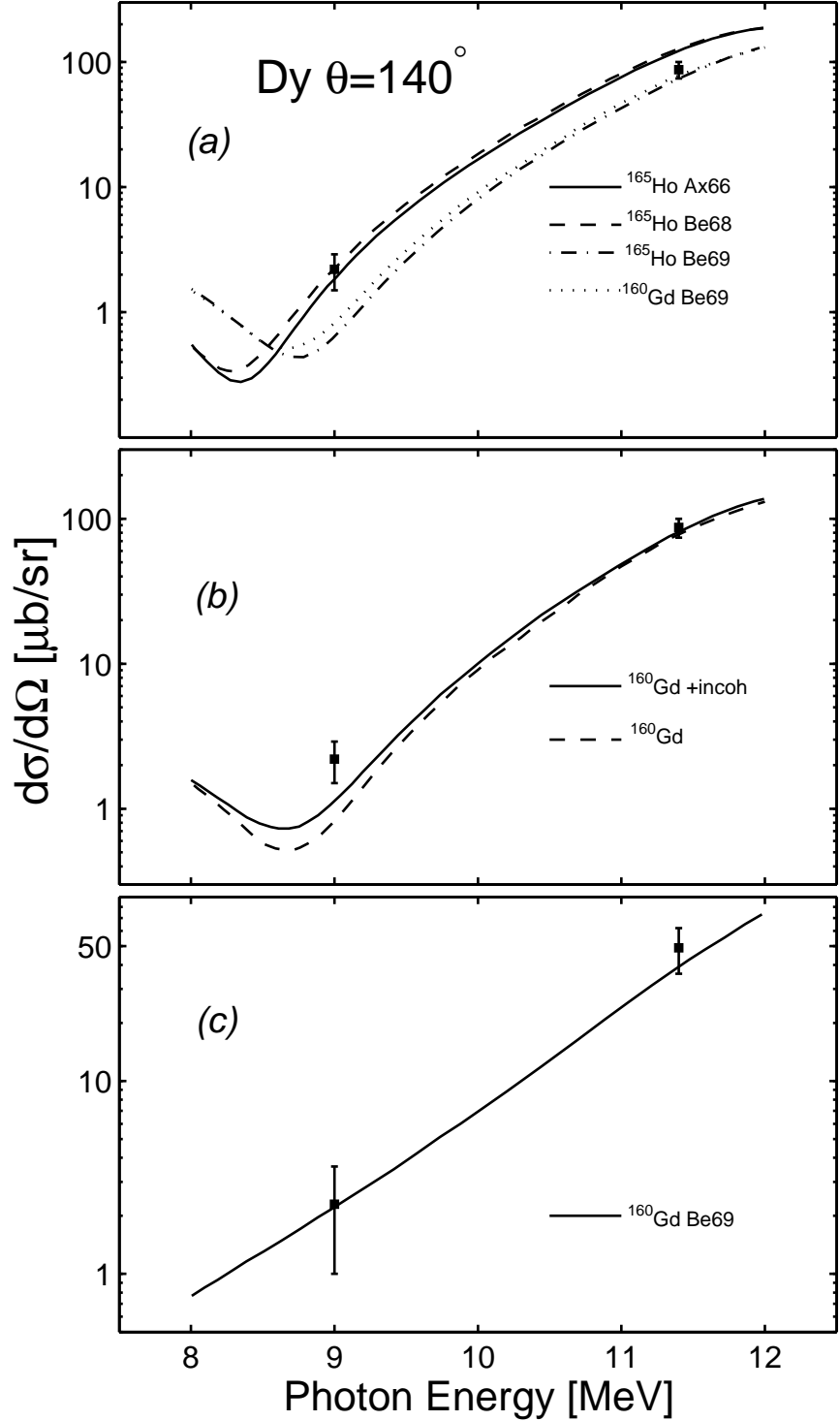}
  \caption{Elastic and inelastic cross sections in Dy, versus calculations with different sets of
           {\em GDR} parameters, taken from neighboring nuclei: {\em (a)} the calculated curves include
           only the elastic {\em coherent} contributions of the Dy isotopes. {\em (b) } The curves are 
           based on $^{160}$Gd parameters. The solid line includes both the {\em coherent} and 
           {\em incoherent} contributions. {\em (c) } Inelastic Raman cross sections. The curve is the 
            result of the Raman contributions
           of $^{162,163,164}$Dy whose first excited states are at $\approx$77 keV.}
  \label{Dyres}
\end{figure}

The analysis of the Dy cross sections is impeded by two factors: \textit{ i)}\ the natural Dy target includes at least 5
isotopes with non-negligible abundances (Table~\ref{Table5}), and \textit{ ii)}\ there are no measurements of
the {\em GDR} parameters for this element. Thus, we tried parameters from the neighbor nuclei of $^{165}$Ho
(Ax66, Be68,Be69) and $^{160}$Gd (Be69).
Results of these calculations are shown in Fig.~\ref{Dyres}a where only the coherent contribution is considered. The
sets of {\em GDR} split in two groups, one giving good agreement at 9.0 MeV and overestimating the 11.4 MeV result, and
one underestimating both results. Two of the isotopes appearing in Table~\ref{Table5} have ground state spins
$I_0 = \frac{5}{2}$
so an incoherent contribution proportional to their relative abundances was added. The best agreement is obtained with
the $^{160}$Gd {\em GDR} set as shown in Fig.~\ref{Dyres}b. The inclusion of the incoherent contribution brings the
calculation at 11.4 MeV in perfect agreement with the experiment, while at 9.0 MeV the discrepancy is markedly reduced.
Therefore we conclude that the unknown {\em GDR} parameters for natural Dy has to be very close to those of
$^{160}$Gd. This conclusion is supported by the calculations of Raman scattering shown in Fig.~\ref{Dyres}c.
Contributions to the Raman scattering cross section were considered to come only from the $^{162,163,164}$Dy isotopes
(with a final excited state at about 77 keV). The contribution of $^{161}$Dy is not included because
its first level energy is at 25.6 keV, being much smaller than the observed Raman energy; $^{160}$Dy
has a too low abundance and was neglected. The good agreement between the data and calculations favors the
Dy {\em GDR}
description by the $^{160}$Gd parameters. On the basis of these parameters we can predict the intrinsic quadrupole
moment $Q_0$. Following Danos~\cite{Da:58}, the ratio $d=a/b$ of the long to short axis of a deformed
nucleus is related to the peak energies of the {\em GDR} by:
\begin{equation}
0.911d+0.089=E_2/E_1
\label{Danos1}
\end{equation}
The intrinsic quadrupole moment is then \cite{Be:69},\cite{Mo:91}:
\begin{equation}
Q_0=\frac{2}{5} Z r_0^2 A^{2/3} \frac{d^2-1}{d^{2/3}}
\label{Danos2}
\end{equation}
with $r_0$=1.2 fm and $E_1$, $E_2$ from $^{160}$Gd {\em GDR} parameters, one obtains for Dy: $Q_0$=7.30~b.
Table~\ref{Table6} summarizes the experimental information on $B(E2)\uparrow$ and static quadrupole moments $Q$ for
various Dy isotopes. The extracted intrinsic $Q_0$ were averaged according to the abundances. The final value for
natural Dy is $Q_0=7.31$~b in excellent agreement with the above prediction.

\subsection{In}
\begin{figure}
  \includegraphics[scale=1.0]{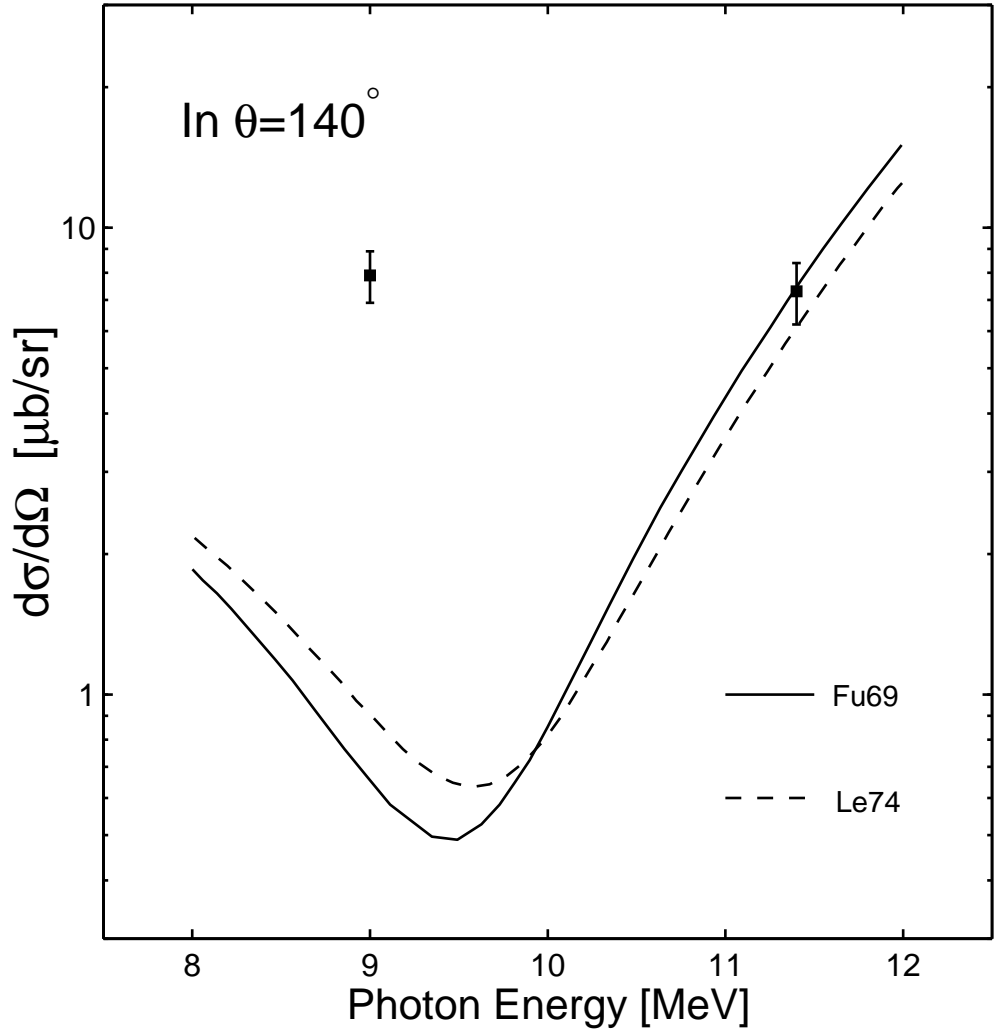}
  \caption{Elastic scattering cross section from In at 140$^\circ$, at 9.0 and 11.4 MeV.}
  \label{Inres}
\end{figure}

Natural In have two isotopes: 4.3\% $^{113}$In and 95.7\% $^{115}$In. The In results shown in
Fig.~\ref{Inres}, represent a challenge with an unexpected high cross section at 9.0
MeV.  An excellent agreement between the measured and the calculated cross section is obtained at
11.4 MeV using the {\em GDR} parameters of Fu69 \cite{Fu:69}. At 9.0 MeV however, the measured value
(Table~\ref{Table3}) is $\approx12$ times higher than the calculated one. This huge departure
can be explained by the resonance excitation of an isolated single compound nuclear level, most
likely in $^{115}$In. The occurrence of such isolated resonance at $\approx 9$ MeV was also observed
in many other nuclei \cite{Ka:94}. 

In this case there is a direct excitation of one or more nuclear levels in the
continuum by the incoming $\gamma$-ray. Such an excitation will be possible when there is a partial overlap between
the incident $\gamma$ energy and its line width with a nuclear level energy and its width. The deexcitation of the
nuclear level back to the ground state will be the measured elastic $\gamma$ scattering. In general, resonance cross
sections (or widths) are subject to strong Porter-Thomas type fluctuations. We shall discuss here only the average 
$\gamma \rightarrow \gamma$ cross section from a nuclear level with spin $I$ \cite{Zu:83}:
\begin{equation}
\bar \sigma_{\gamma \gamma}^I(E_\gamma)  =   \pi^2 \left(\frac\lambda{2\pi}\right)^2 g \eta \left(
\zeta \right) 
\left( \bar \frac{\Gamma_0^2}{ \bar \Gamma D} \right ) 
\end{equation}
where $\bar \Gamma_0$ - the average ground state width (transitions to ground state); $\bar \Gamma$ - the
average total decay width; $D$ - nuclear level spacing obtained from $\rho_I(E)$ the nuclear level density;
$\eta(\zeta)$ - an enhancement function depending on the ratio 
$\zeta = \bar \Gamma_{ex}/\bar \Gamma_0 = (\bar \Gamma - \bar \Gamma_0)/\bar \Gamma_0$ where $\bar \Gamma_{ex}$ is
the average total $\gamma$ width for transitions to the excited states; $g$ - the statistical factor $(2I+1)/(2I_0+1)$ for
transitions from an excited state $I$ to the ground state $I_0$; $\lambda$ - the wavelength of the scattered radiation
of energy $E_\gamma$.
The function $\eta(\zeta)$ changes from 1 for $\zeta=0$ (transitions to the ground state only) to 3 for $\zeta=\infty$
(no transitions to the ground state at all).  

$\bar \Gamma_0$ is obtained from the photoabsorption cross section, described by the
{\em GDR} parameters in Table~\ref{Table2}: 
\begin{eqnarray}
\sigma_{{\rm ph}}(E_\gamma)  & = & \sigma_{{\rm GDR}} \frac{\Gamma_{{\rm GDR}}^2 E_\gamma^2}
{(E_{{\rm GDR}}^2-E_\gamma^2)^2 + \Gamma_{{\rm GDR}}^2 E_\gamma^2} \nonumber   \\
\sigma_{{\rm ph}}(E_\gamma) & = & 3 \pi^2 \left( \frac{\lambda}{2\pi} \right)^2 
\frac{\bar \Gamma_0(E_\gamma)}{D(E_\gamma)}                                   \nonumber                      
\end{eqnarray}

For $\bar \Gamma$ we took the experimetal value~\cite{Ly:68} 81 meV, measured at neutron separation energy in thermal
capture.

The average differential cross section will be given by:
\begin{equation}
\frac{{\rm d}\bar \sigma_{\gamma \gamma}(\theta)} {{\rm d}\Omega} = \sum_I \frac{\bar \sigma_{\gamma \gamma}^I}{4\pi} 
\left(1+A_{22}^I P_2(\cos \theta)\right)
\end{equation}
where the $A_{22}^I$ coefficients for $E1$ transitions in the cascades $I_0\rightarrow I\rightarrow I_0$ with
$I_0=\frac92$ (the ground state for $^{115}$In) and $I=\frac72,\ \frac92,\ \frac{11}2$ are 0.02333, 0.19394 and 0.08273 
respectively.

The level density $\rho_I(E)$ was evaluated with a back shifted formula. The parameters $a=14.086$ MeV$^{-1}$ and
$\delta=-0.63$ MeV were taken from the $RPIL$ library~\cite{So:73}. The two sets of $^{115}$In {\em GDR} parameters in Table
\ref{Table2} give at $E_\gamma=9.0$ MeV (taking $\eta(\zeta)=1$) ${\rm d}\bar \sigma_{\gamma \gamma}(\theta=140^\circ)/
{\rm d}\Omega = 8.4$ and 8.7 $\mu$b/sr respectively, in fair agreement with the measured value $7.9\pm1.1\ \mu$b/sr.

\section{Conclusions}
The elastic scattering cross sections in Au are nicely reproduced using Be86 {\em GDR} parameter set available 
in the literature. The observation of weak Raman transitions are viewed as an evidence for the
occurence of a slight deformation in $^{197}$Au. Qualitative and some quantitative agreement with these Raman transitions
is obtained when a two peaks {\em GDR} with small energy difference is enforced.

In Dy both the elastic and Raman intensities were found to agree when the {\em GDR} parameters of the neighboring 
$^{160}$Gd nucleus were employed. Therefore the natural Dy {\em GDR} parameters are expected to be very close to those
of $^{160}$Gd.

At 9.0 MeV in In an isolated resonance was excited in $\gamma \gamma$ scattering. The measured cross section agrees with
calculations based on the statistical model of the nucleus. At 11.4 MeV the character of the nuclear excitation changes and
becomes a collective {\em GDR} type. At this energy agreement is obtained with the Fu69 parameters.

\newpage

\begin{table}
\caption{Amplitudes (in units of $r_0$) for 9.0 MeV photons elastically scattered from Au at 
         $\theta=140^\circ$.}
\label{Table1}
\begin{ruledtabular}
\begin{tabular*}{\hsize}{c@{\extracolsep{0ptplus1fil}}
                 r@{\extracolsep{0pt}}@{}l@{\extracolsep{0ptplus1fil}}
                 r@{\extracolsep{0pt}}@{}l}
Amplitude  & \multicolumn{2}{c}{$\parallel$}  & \multicolumn{2}{c}{$\perp$}                  \\
\colrule 
{\em T}  & $+1$  & $.331\times10^{-2}$          & $-1$  & $.738\times10^{-2}$          \\
{\em NR} & $-(1$ & $.580+i0.496)\times10^{-2}$  & $+(2$ & $.063+i0.648)\times10^{-2}$  \\
{\em D}  & $+(0$ & $.252+i0.274)\times10^{-2}$  & $-(0$ & $.186+i0.222)\times10^{-2}$  \\
{\em R}  & $-0$  & $.00095\times10^{-2}$        & $+0$  & $.0012\times10^{-2}$         \\
\end{tabular*}
\end{ruledtabular}
\end{table}

\begin{table*}
\caption{Sets of {\em GDR} parameters used in the present experiment. The original experiments are referenced; the
actual parmeters were taken from the Lorentzian fits of Dietrich and Berman \cite{Di:88}.}
\label{Table2}
\begin{ruledtabular}
\begin{tabular*}{\hsize}{c@{\extracolsep{0ptplus1fil}}c@{\extracolsep{0ptplus1fil}}c@{\extracolsep{0ptplus1fil}}c@{\extracolsep{0ptplus1fil}}
c@{\extracolsep{0ptplus1fil}}c@{\extracolsep{0ptplus1fil}}c@{\extracolsep{0ptplus1fil}}c@{\extracolsep{0ptplus1fil}}c}
Ref.  & Symbol  & $E_1$ [MeV]  &   $\sigma_1$ [mb] & $\Gamma_1$ [MeV]  & $E_2$ [MeV]  &   $\sigma_2$ [mb] &
$\Gamma_2$ [MeV]  & Nucleus \\
\colrule
\cite{Fu:62} & Fu62 & 13.82 & 560 & 3.84 &&&& $^{197}$Au \\
\cite{Ve:70} & Ve70 & 13.72 & 541 & 4.61 &&&& $^{197}$Au \\
\cite{Be:86} & Be86 & 13.73 & 502 & 4.76 &&&& $^{197}$Au \\
\cite{So:73} & So73\footnote{Lorentzian parameters from Varlamov data in the RIPL library.} & 13.60 & 590 & 4.50
                    &&&& $^{197}$Au \\
Present\footnote{Two Lorentzian fit to the combined data of Ve70 and Be86 performed in this work.}
                  & & 13.70 & 260 & 3.0  & 13.90 & 290 & 5.3 & $^{197}$Au \\
\colrule
\cite{Ax:66} & Ax66 & 12.02 & 238 & 2.35 & 15.59 & 308 & 4.85 & $^{165}$Ho \\
\cite{Be:69} & Be69 & 12.28 & 214 & 2.57 & 15.78 & 246 & 5.00 & $^{165}$Ho \\
\cite{Be:68} & Be68 & 12.01 & 239 & 2.52 & 15.59 & 291 & 5.12 & $^{165}$Ho \\
\cite{Be:69} & Be69 & 12.23 & 215 & 2.77 & 15.96 & 233 & 5.28 & $^{160}$Gd \\
\colrule
\cite{Fu:69} & Fu69 & 15.63 & 266 & 5.24 &&&& $^{115}$In \\
\cite{Le:74} & Le74 & 15.72 & 247 & 5.60 &&&& $^{115}$In \\
\end{tabular*}
\end{ruledtabular}
\end{table*}

\begin{table}
\caption{Differential cross sections d$\sigma(\theta=140^\circ)/d\Omega$ in $\mu$b/sr, measured in the present
experiment.}
\label{Table3}
\begin{ruledtabular}
\begin{tabular}{c@{\extracolsep{0ptplus1fil}}c@{\extracolsep{0ptplus1fil}}
               r@{\extracolsep{0pt}}@{}l@{${}\pm{}$}r@{}l@{\extracolsep{0ptplus1fil}}
               r@{\extracolsep{0pt}}@{}l@{${}\pm{}$}r@{}l}
Target  &   & \multicolumn{4}{c}{11.4 MeV}  &   \multicolumn{4}{c}{9} MeV \\
\colrule
Au\footnote{Additionally, we are using in Fig.~\ref{XSfig} two cross sections measured separately with
a Cr(n,$\gamma$) photon source: $2.2\pm0.3$ at 8.88 MeV and $7.0\pm0.8$ at 9.72 MeV.}
   & Elastic & 116&  & 17&   & 3& .0 & 0 & .9 \\
Dy & Elastic &  87&  & 13&   & 2& .2 & 0 & .7 \\
   & Raman   &  49&  & 13&   & 2& .3 & 1 & .3 \\
In & Elastic &   7&.1&  1&.1 & 7& .9 & 1 & .0 \\
\end{tabular}
\end{ruledtabular}
\end{table}

\begin{table}
\caption{Measured and calculated inelastic differential cross sections in $\mu$b/sr
leading to low lying levels in $^{197}$Au.}
\label{Table4}
\begin{ruledtabular}
\begin{tabular}{c@{\extracolsep{0ptplus1fil}}c@{\extracolsep{0ptplus1fil}}d
                r@{\extracolsep{0pt}}@{}l@{${}\pm{}$}r@{}l
                 @{\extracolsep{0ptplus1fil}}d}
  &  &  & \multicolumn{4}{c}{Experimental}        \\
$K_0$  & Level Spin   &\multicolumn{1}{c}{Level Energy [keV]}   & \multicolumn{4}{c}{cross section} & 
\multicolumn{1}{c}{Raman}  \\
\colrule
$\frac{1}{2}$  & $\frac{1}{2}^+$ & 77.351   & 1&.7& 1&.2  &  1.8       \\
$\frac{1}{2}$  & $\frac{3}{2}^+$ & 268.786  & 1&.1& 0&.9  &  1.8       \\
$\frac{1}{2}$  & $\frac{5}{2}^+$ & 502.5    & 0&.2& 0&.8  &  0.8       \\
$\frac{1}{2}$  & $\frac{7}{2}^+$ & 736.7\footnote{Not observed in the present experiment.} &
    \multicolumn{4}{c}{?}  &                                 4.6       \\
\colrule
$\frac{3}{2}$ & $\frac{3}{2}^+$ & 0        & 3&.0& 0&.9\footnote{Elastic cross section from Table~\ref{Table3}. 
Most of it is the coherent part not related to the Raman scattering.} &   
1.8\footnote{Calculated incoherent contribution to the elastic scattering.}      \\
$\frac{3}{2}$ & $\frac{5}{2}^+$  & 278.99   & 0&.3& 0&.7  &   4.6          \\
$\frac{3}{2}$ & $\frac{7}{2}^+$  & 547.5    & 1&.3& 0&.9  &   3.5        
\end{tabular}
\end{ruledtabular}
\end{table}

\begin{table}
\caption{Natural abundance, ground and first excited state energies and spin of stable Dy isotopes.}
\label{Table5}
\begin{ruledtabular}
\begin{tabular*}{\hsize}{c@{\extracolsep{0ptplus1fil}}d@{\extracolsep{0ptplus1fil}}c@{\extracolsep{0ptplus1fil}}
d@{\extracolsep{0ptplus1fil}}c}
     &                                       & I$_0$        & \multicolumn{1}{c}{E [keV]}     & I$_f$   \\
A    &  \multicolumn{1}{c}{Abundance [\%]}   & ground state & \multicolumn{1}{c}{first level} & first level \\
\colrule
160 & 2.3  & $0^+$            &  86.8  &  $2^+$            \\
161 & 18.9 & $\frac{5}{2}^+$  &  25.6  &  $\frac{5}{2}^-$  \\
162 & 25.5 & $0^+$            &  80.7  &  $2^+$            \\
163 & 24.9 & $\frac{5}{2}^-$  &  73.3  &  $\frac{7}{2}^-$  \\
164 & 28.2 & $0^+$            &  73.4  &  $2^+$            \\
\end{tabular*}
\end{ruledtabular}
\end{table}

\begin{table}
\caption{Derivation of the intrinsic quadrupole moment $Q_0$ from the mesured $B(E2)\uparrow$ values (even masses)
and static quadrupole moments $Q$ (odd masses) for Dy isotopes.}
\label{Table6}
\begin{ruledtabular}
\begin{tabular*}{\hsize}{c@{\extracolsep{0ptplus1fil}}c@{\extracolsep{0ptplus1fil}}c@{\extracolsep{0ptplus1fil}}
c@{\extracolsep{0ptplus1fil}}c}
A    &  $B(E2)\uparrow$\footnote{data taken from \cite{Ra:87}}\ [e$^2$b$^2$]   &
$Q_0$\footnote{$Q_0=\left[16 \pi /5 \times B(E2)\uparrow /e^2\right]^{1/2}$, from \cite{Bo:75} Eq. 4-68.}\ [b] &
$Q$\footnote{data taken from \cite{Ra:89}.}\ [b] &
$Q_0$\footnote{$Q_0=(I_0+1)(2I_0+3)/(I_0(2I_0-1)) \times Q$, from \cite{Bo:75} Eq. 4-70.}\ [b] \\
\colrule
160 & 5.06  & 7.13  &         &              \\
161 &       &       &  2.494\footnote{average of three values.}  &  6.98  \\
162 & 5.28  & 7.28  &         &              \\
163 &       &       &  2.648  &  7.41  \\
164 & 5.6   & 7.5   &         &              \\
\end{tabular*}
\end{ruledtabular}
\end{table}

\newpage

\begin{figure}
   \includegraphics[scale=1.0]{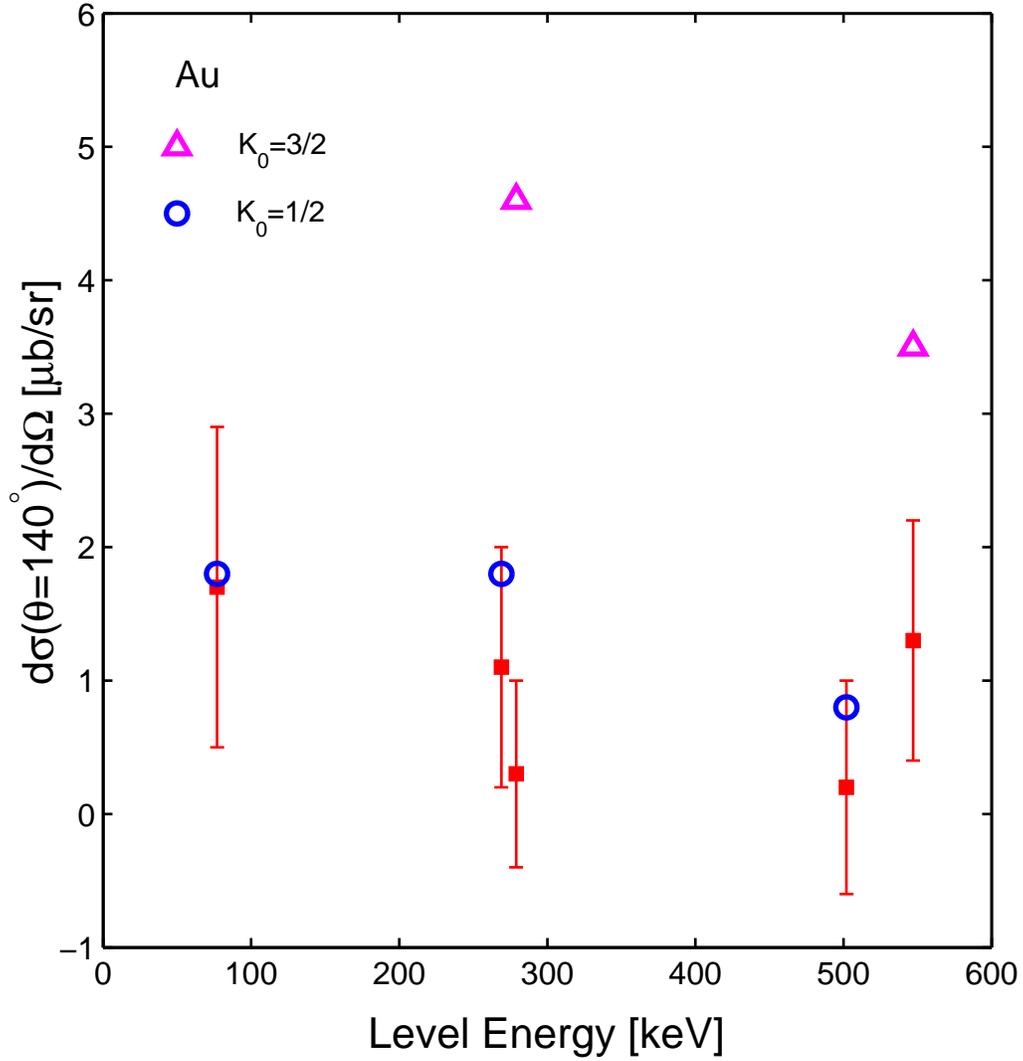}
   \caption{Raman inelastic scattering cross sections (squares) and theoretical calculations,
            $K_0=\frac{1}{2}$ circles and $K_0=\frac{3}{2}$ triangles,
            in $^{197}$Au at 9.0 MeV, as a function of the excitation energies of the final states.}
   \label{RamanTheory}
\end{figure}

\end{document}